\documentclass[a4paper]{article}

\usepackage{INTERSPEECH2021}
\usepackage{hyperref}

\title{Location, Location: Enhancing the Evaluation of Text-to-Speech synthesis using the Rapid Prosody Transcription Paradigm}
\name{Elijah Gutierrez$^1$, Pilar Oplustil-Gallegos$^2$, Catherine Lai$^{1,2}$}
\address{
  $^1$Linguistics and English Language, University of Edinburgh, United Kingdom\\
  $^2$The Centre for Speech Technology Research, University of Edinburgh, United Kingdom}
\email{s1740779@sms.ed.ac.uk, p.s.oplustil-gallegos@sms.ed.ac.uk,  c.lai@ed.ac.uk}

\begin{document}

\maketitle
\begin{abstract}
Text-to-Speech synthesis systems are generally evaluated using Mean Opinion Score (MOS) tests, where listeners score samples of synthetic speech on a Likert scale.  A major drawback of MOS tests is that they only offer a general measure of overall quality—i.e., the naturalness of an utterance—and so cannot tell us where exactly synthesis errors occur.  This can make evaluation of the appropriateness of prosodic variation within utterances inconclusive.     
To address this, we propose a novel evaluation method based on the Rapid Prosody Transcription paradigm. This allows listeners to mark the locations of errors in an utterance in real-time, providing a probabilistic representation of the perceptual errors that occur in the synthetic signal.   
We conduct experiments that confirm that the fine-grained evaluation can be mapped to system rankings of standard MOS tests, but the error marking gives a much more comprehensive assessment of synthesized prosody.  
 In particular, for standard audiobook test set samples, we see that error marks consistently cluster around words at major prosodic boundaries indicated by punctuation.  However, for question-answer based stimuli, where we control information structure, we see differences emerge in the ability of neural TTS systems to generate context-appropriate prosodic prominence.  
\end{abstract}
\noindent\textbf{Index Terms}: Speech Synthesis, TTS, TTS Evaluation, MOS, Prosody, Rapid Prosody Transcription, Speech Perception

\section{Introduction}
Modern text-to-speech (TTS) systems have attained a level of naturalness that is approaching human parity for isolated utterances \cite{malisz2019modern}. This progress is in large part due to the rise of neural network based machine learning methods, which have drastically improved the overall quality of synthetic speech and enabled researchers to focus more attention on generating natural sounding prosodic variation. In recent years, there has been substantial research on achieving fine-grained control over synthetic prosody  \cite{Watts2015control,Hodari2020perception,lee2019robust}. New prosodic control mechanisms have allowed TTS systems to produce more variable and expressive speech \cite{Hodari2019ssw}. However, there has been relatively little work determining whether the prosody that is assigned to an utterance is actually licensed by a given context \cite{clark2019ssw,Tyagi2019}, and it is not clear whether current subjective evaluation methods, such as Mean Opinion Score (MOS) tests, provide enough information to determine the contextual appropriateness \cite{Wagner2019}. 

The appropriatess of utterance prosody--which broadly includes pitch, energy, timing and other suprasegmental characteristics of speech--can vary greatly depending on context. In fact, prosodic differences can help disambiguate many aspects of discourse and dialogue structure \cite{lai2020integrating,kleinhans2017using,shriberg1998can,tran2020neural,gravano2011turn}.
Many studies have also shown the close relationship between context-induced expectations about the prosodic form of utterances and information structural notions like newness and givenness \cite{Steedman2000,calhoun2010centrality,Breen2010}.
Incorporating discourse relations has been shown to improve the perceived naturalness of synthesized speech \cite{aubin2019improving}, while incorporating 
information structure into generated speech has been shown to improve naturalness of automated task oriented dialogues \cite{white2010generating}. As neural TTS models continue to improve in their ability to generate variable prosody, it is important to note that not all variation is appropriate in all contexts and increased variation within an utterance is not always perceived as natural \cite{Watts2015control}. 

In order to evalute how and where TTS systems are really improving in terms of prosody, we need methods that give us a clearer view of what sort of prosodic patterns they generate, and how their appropriateness changes with context.
To do this, we propose a new evaluation method that augments traditional MOS-based listening tests with finer-grained  error annotations.  Specifically, we draw on the Rapid Prosody Transcription (RPT) framework \cite{mo2008naive,Cole2016} to obtain information about the location of perceived errors in the prosody of synthesized speech. 
In RPT, non-expert listeners mark the presence of prosodic phenomena (e.g. prominence or boundary placement) in real time.  
This approach allows us to  more precisely identify contextual/linguistic sources of prosodic errors.

Much of the current work on TTS in context has focused on monologue or narrative style generation, where information structural relationships are generally unclear \cite{clark2019ssw,Tyagi2019}, and prosodic expectations may not be strong. To address this,  we created a schema for generating question-answer pairs with well defined information structure, which in turn project clear prosodic expectations for synthesized answers. Combined with word-level error annotation, this allows us to identify cases of contextually inappropriate prosodic variation.

In the following, we show that there is a strong negative correlation between measures based on error marking and MOS, from which we can recreate MOS based system rankings. Moreover, our question-answer stimuli can be used to induce stronger expectations about prosody than classic audiobook style test utterances, and so better highlights differences in the system prosodies. 
In general, inspection of the distibution of errors across systems for specific stimuli can lead to better understanding of the sources of system differences, which may otherwise be obscured by MOS alone.

\section{Background}

TTS researchers have developed a wide range of methods to evaluate the quality of synthetic speech \cite{Wagner2019}. However, subjective methods are still considered to be the  gold standard in TTS evaluation. These generally involve asking listeners to rate speech samples on a specified dimension, usually naturalness (i.e., how 'humanlike' synthesized speech sounds). The most commonly used subjective evaluation type is the Mean Opinion Score (MOS) test: listeners are presented with a synthetic stimulus and asked about their overall impression of it, scoring the stimulus on a (usually 5-point) Likert scale \cite{van1995quality}.

Some of the advantages of MOS tests are that they are straightforward to set up, they are quicker and less cognitively taxing than ranking tasks like MUSHRA, and MOS test design choices have been well extensively investigated \cite{clark2019ssw,wester2015listeners}. 
Recent work has expanded to evaluation beyond the single sentence \cite{clark2019ssw,Tyagi2019}. 
However, these studies generally focus on holistic evaluations over multi-utterances segments, rather than the parts of utterances that may change listener perception.  
Meanwhile, the relationship between linguistic context and sub-utterance prosody has been extensively studied, particularly in English, in terms of information structure \cite{calhoun2010centrality,Breen2010,roettger2019mapping}, i.e. how information is organised in an utterance. This is  usually cast in terms of given/new information (similarly topic/focus). These constraints are usually demonstrated using question-answer constructions:  For example, `ALEX ate the brownies' is an appropriate answer to `Who ate the brownies?' because `Alex' is the new information, while `Alex ate the BROWNIES' is infelicitous because 'brownies' is contextually given. Thus, use of stimuli with clear information structure provides a precise way of probing whether prosody is context appropriate or not. 

Though Information Structure theory can give us an idea of prosodic expectations, prosody perception is known for high inter-listener variability \cite{mo2008naive,Cole2016} even with expert training \cite{syrdal2000tobi}. So, the fact that the RPT framework was specifically developed to capture variability in prosody perception from non-expert listeners makes it a natural choice for exploring the perception of synthesized speech. In RPT, the perceptual salience of a prosodic features (e.g. prominence) is determined by the number of listeners who mark a specific segment (e.g. word) with that feature.  
Because RPT is a task that is conducted in real-time, responses are more sensitive to subtle local changes in quality than offline ones \cite{van1995quality,ralston1991comprehension}.

While RPT has been extensively used to study perception of prosody in human speech \cite{Cole2015}, there has been little empirical work investigating within utterance prosody for TTS. The closest related work is Edlund et al.'s Audience Response System \cite{edlund2015audience}, where a group of listeners judged a synthetic sample simultaneously, pressing a button whenever they hear ‘oddities’. This was used to identify common error regions and find the average response latency of listeners when marking errors. However, the definition of a perceptual error was left (deliberately) unclear.  Edlund et al. used a single long-form stimulus of about 3 minutes in length. In contrast, our study compares different types of stimuli across multiple TTS systems and focuses on prosody.

\section{Experimental Setup}

\subsection{Experiments and Hypotheses}

We perform three listening tests to probe the usefulness of our proposed evaluation method.
Experiment condition 1 (E1) is a standard MOS test, while Experiment condition 2 (E2) is a MOS test augmented with the RPT-based error marking task. We compare the results of these two tests to see whether orienting the evaluation to prosody and adding the error marking task affects MOS results.  

In E1 and E2 listeners rate single utterances taken from  the widely used LibriTTS test set \cite{zen2019libritts}.
In Experiment condition 3 (E3), listeners completed the augmented MOS test on question-answer stimuli designed to evoke specific information structural expectations. The goal here was to determine if the error marking would bring out listener expectations about utterance prosody, and hence allow us to distinguish between the appropriateness of prosodic renditions more precisely. This also allows us compare the types of error marks between the two types of test data (audiobook vs dialogue).

\subsection{TTS systems}
In each of the 3 listening tests, we compare three TTS systems: the Festival \cite{clark2007multisyn}, Ophelia \cite{ophelia}, and FastPitch \cite{lancucki2020fastpitch}. 

\emph{Festival} is a standard toolkit for building synthetic voices with unit selection. For these experiments, the ‘SLT’ voice distributed by FestVox was used, i.e. a female voice with a General American accent, built from the Arctic A corpus. We note that this voice is far from the current state-of-the-art in TTS, and so we use it as a baseline to see if listeners would ignore other signal naturalness issues  when asked to attend to prosodic errors. 

We use two neural TTS models as representative of the current state of the art in TTS. These were both trained on the Linda Johnson (LJ) Speech dataset \cite{LJ}, which consists of 13,100 recorded utterances from 7 non-fiction books. \emph{Ophelia} models were trained using the default recipe (500 epochs for Text2Mel, 250 epoch for SSRN). \emph{FastPitch} stimuli were synthesised using character (rather than phone) inputs via a pre-trained sequence-to-sequence model that was trained for 1000 epochs. 

\subsection{Stimuli}

For E1 and E2, 30 sentences were sampled randomly from the evaluation set of the LibriTTS corpus \cite{zen2019libritts}, a popular audiobook corpus specially designed for TTS research. The maximum stimulus length was controlled to be 15 words to mitigate listener boredom and fatigue. 

For E3, contexts and stimuli were generated in a similar manner to those used by \cite{Breen2010} for their study of the acoustic correlates of information structure. We used a template-based approach, involving simple \emph{Subject Verb Object} sentences, generating two types of question-answer pairs:
\begin{itemize}
    \item Informational Focus: SVO\\
            e.g., Q: \emph{What did Mary eat?}\\
            \phantom{e.g.,} A: \emph{Mary ate the cake.}
    \item Corrective Focus: No, SVO \\
            e.g. Q: \emph{Did Mary buy the cookies?}\\ 
            \phantom{e.g.,} A: \emph{No, John bought the cookies.}
\end{itemize}

Questions were generated to change which constituent represented the new information/correction in the answer, which in English determines the appropriate prominence placement in the response stimuli. 
We created 10 stimuli per prominence position. Since there were two stimulus structures, this resulted in 10×3×2=60 stimuli in total.  

\subsection{Evaluation Tasks}
The experiments were designed and distributed remotely using a customized version of the Language Markup and Experimental Design Software (LMEDS) \cite{mahrt2016lmeds}. Each stimulus was presented on its own page as follows. 

For the standard MOS test (E1), a transcript of the audio stimulus was presented with a ‘Play’ button. Participants were asked to answer the question \emph{`How natural does the speaker sound?'} on a 5-point Likert scale via a scale slider (MOS).

The augmented MOS tests (E2, E3) included the additional RPT-based error marking task, the MOS slider, and a further error type survey. The error marking task appeared first: participants were asked to listen to the stimulus and to click on any words in the transcript where the intonation did not sound correct (possibly none), highlighting them in red. For E3, participants were told to read the context question before marking errors on the answer stimulus. Participants were allowed to replay the stimulus up to 3 times and change their error marks. The MOS slider was positioned after the error marking task, rating  \emph{'How natural is the speaker's intonation?'} on a 5-point scale (PMOS). 
Finally, participants were asked to select which error types they noticed out of: ‘Abrupt change in pitch’, ‘Awkward pause’, ‘Unexpected intonation’ and ‘Lacking intonation’.  
These choices were based on our initial impressions of potentially common errors. 
Participants also had access to an ‘Other’ box to enter additional comments or a custom response.
In E2 and E3, participants were initially shown 3 examples of stimuli with prosodic errors along with an explanations of why they were considered odd or unnatural. Once this familiarisation phase was complete, participants moved on to the main task. In all three experiments, the audio stimuli were presented in a random order.

\subsection{Participants and Groups}
English-speaking participants were recruited with the crowd sourcing platform Prolific Academic.\footnote{\url{https://www.prolific.co}}  
Each participant was paid £2 for their participation in the study.

Participants were assigned a random group via Prolific and directed to a listening test based on a Latin square design for each evaluation condition (E1, E2: 3 groups of 10, E3: 6 groups of 10). Participants evaluated stimuli from every system, but did not evaluate the same text stimulus more than once.

After consenting to participate in the study, participants were instructed to wear headphones for best audio quality, to ensure they had a stable connection to the server, and to focus their attention on the evaluation task. A brief explanation of what was meant by intonation was also given for E2 and E3. Each participant rated 30 stimuli via the LMEDS interface described above. The standard MOS (E1) test took 8 minutes to complete on average, while the augmented MOS tests (E2, E3) took 15 minutes.\footnote{Further details/stimuli:  \url{http://sweb.inf.ed.ac.uk/clai/tts-rpt}}

\section{Results}

\begin{figure}[t]
    \centering
    \includegraphics[scale=0.55]{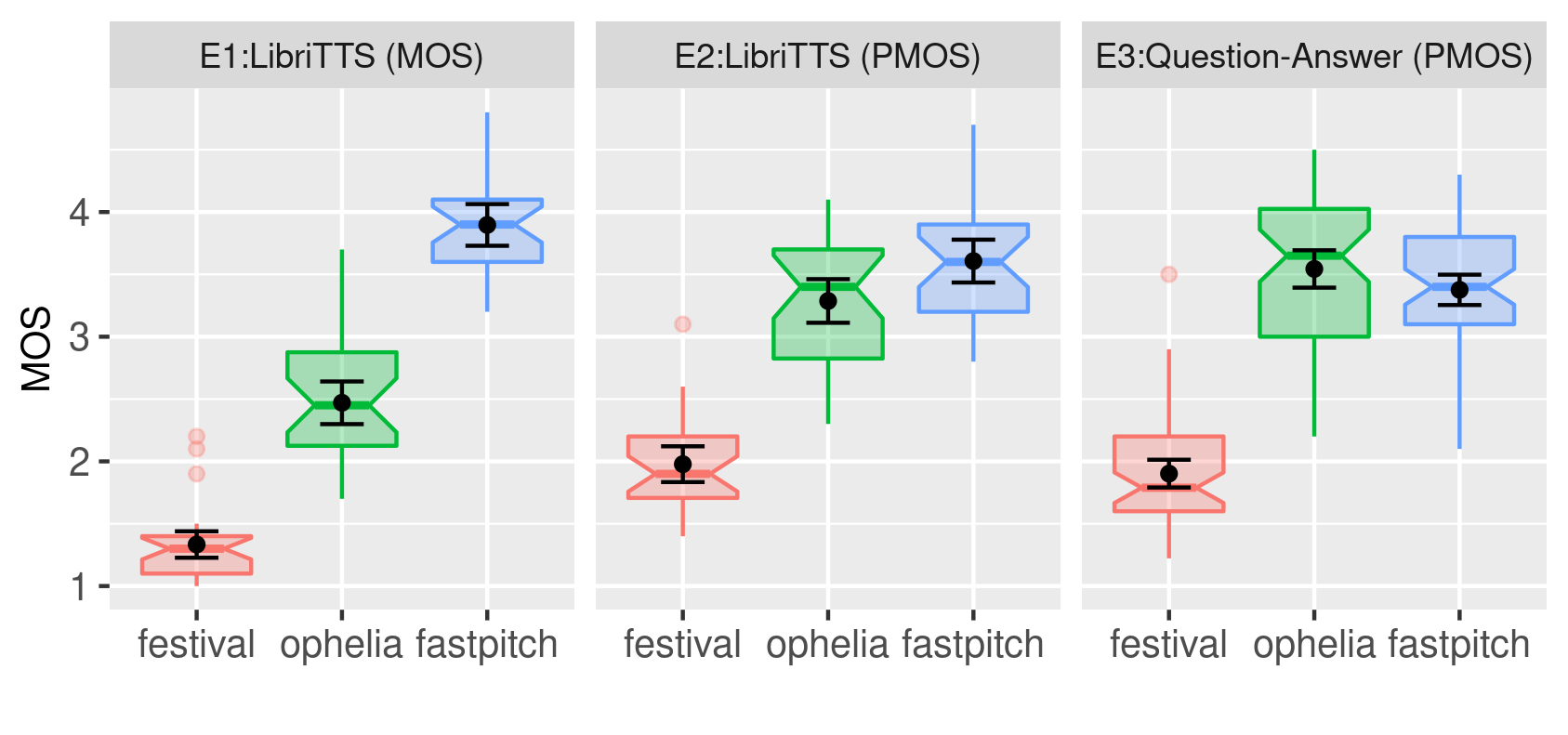}
    \caption{Distribution of Mean Opinion Scores  per experiment (boxplots and means with 95\% confidence intervals in black).} 
    \label{fig:grand-mos-boxplot}
    \vspace{-5mm}
\end{figure}

\begin{table}[ht]
\centering
\caption{Mean / IQR per stimulus mean MOS}
\begin{tabular}{lccc}
  \hline
System & E1 (MOS) & E2 (PMOS) & E3 (PMOS) \\
  \hline
 Festival & 1.33 / 0.30 & 1.98 / 0.49 & 1.90 / 0.60  \\ 
 Ophelia & 2.47 / \emph{0.75} & 3.29 / \emph{0.88} & \textbf{3.54} / \emph{1.03} \\ 
  FastPitch & \textbf{3.90} / 0.50 & \textbf{3.61} / 0.70 & 3.38 / 0.70 \\ 
   \hline
\end{tabular}
\label{tab:mean-iqr-mos}
\end{table}

Figure~\ref{fig:grand-mos-boxplot} shows the distribution of mean Likert scale ratings per stimuli for the three experimental conditions. For E1 this is the classic 'naturalness' (MOS), while for E2 and E3 this is a prosodic naturalness rating (PMOS).  Comparing the results for E1 and E2 (LibriTTS), we see that the MOS and PMOS scores show the same overall ranking of systems. However, the absolute difference between the system means is reduced in E2, with a marked increase for the  scores for Festival and Ophelia. 
Table~\ref{tab:mean-iqr-mos} shows the means and Interquartile ranges (IQR) for the 3 tests (IQR is reported as a measure of dispersion for consistency instead of standard deviation as system distributions were skewed). 
The overall mean MOS is significantly different for all systems in E1 (paired t-test, $p<0.01$ with Bonferroni correction), resulting in the ranking Fastpitch $>$ Ophelia $>$ Festival. However, in the PMOS conditions (E2, E3), the difference between FastPitch and Ophelia is no longer significant at the same level (i.e, $p > 0.01$).  Ratings of Ophelia-produced stimuli were the most variable for all conditions, with the greatest dispersion shown for the question-answer condition. 

These distributional differences indicate that shifting participants focus to prosodic errors changed how they rated the stimuli. This also suggests that  lower ratings for Festival and Ophelia in E1 were due to non-prosodic issues. Conversely, the higher ratings for FastPitch are for overall better synthesis quality, but not necessarily for more natural prosodic realization.   
As we shift to test stimuli with clearer prosodic expectations, the gap between systems in terms of prosodic naturalness is reduced and sometimes reversed relative to what we'd expect given only a standard MOS naturalness test. 

\begin{figure}
    \centering
    \includegraphics[scale=0.7]{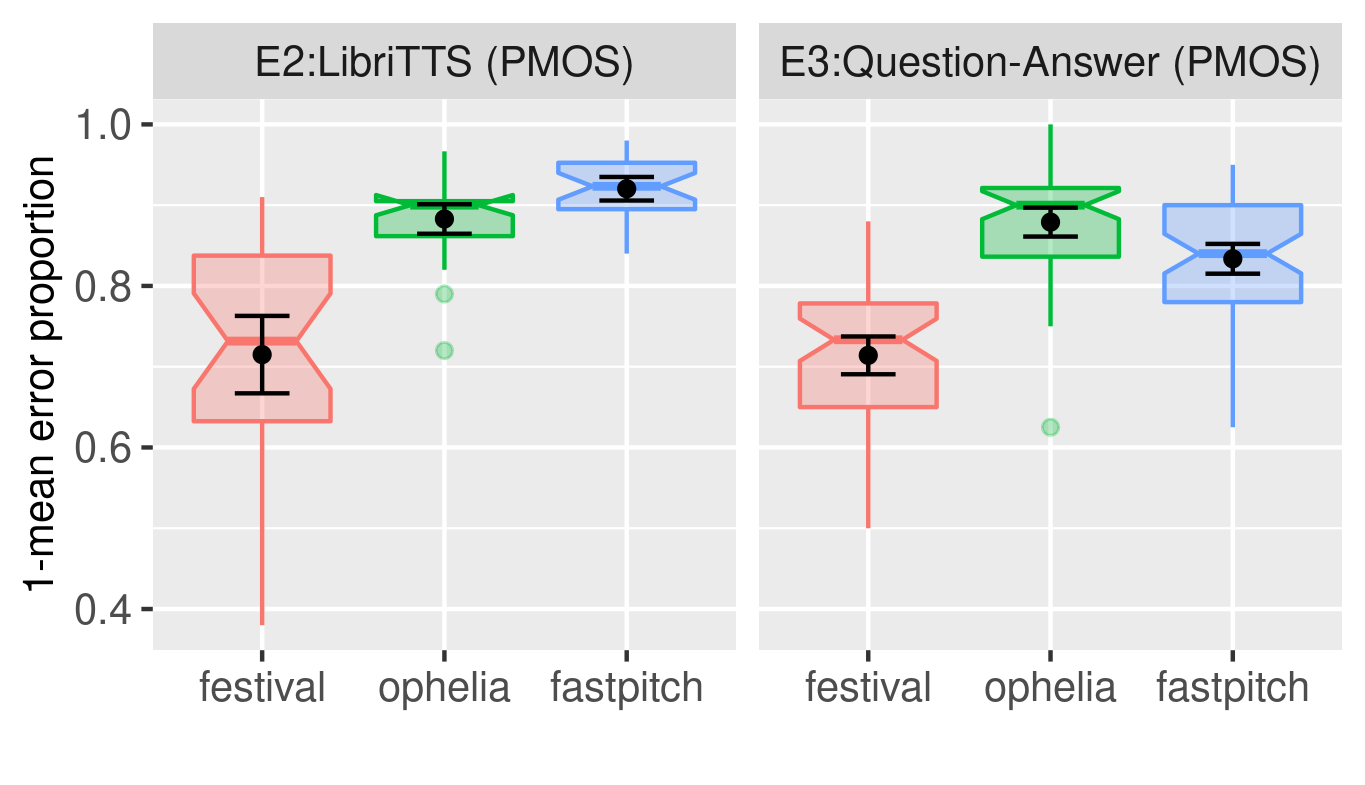}
    \caption{Distribution of mean error rates (per stimulus).}
    \label{fig:grand-mean-pc-error-boxplot}
\vspace{-3mm}
\end{figure}

To see how the error marking task relates to PMOS, we calculated the error marking rate (number errors/number of words) per stimuli and participant. Figure~\ref{fig:grand-mean-pc-error-boxplot} shows the distribution of the mean error rate per stimuli (shown as 1-mean error rate to mirror PMOS ranking). We see that the overall system rankings are the same as that shown in Figure~\ref{fig:grand-mos-boxplot} for PMOS.  Unsurprisingly, the correlation between stimulus PMOS and error rate is strongly negative when we pool data across all conditions (Pearson’s $R=-0.75$).  All differences in mean error rate are significant (paired t-tests, $p<0.01$, Bonferroni correction) except between Ophelia and FastPitch in E2, i.e. when we look at the word level errors in for question-answer stimuli Ophelia performs significantly better than FastPitch. This indicates that the fine-grained evaluation has better ability to differentiate the system prosody when prosodic expectations are designed to be stronger (E3), but this difference may not be apparent for classic narrative style test sets. 

\begin{figure}
    \centering
    \includegraphics[scale=0.7]{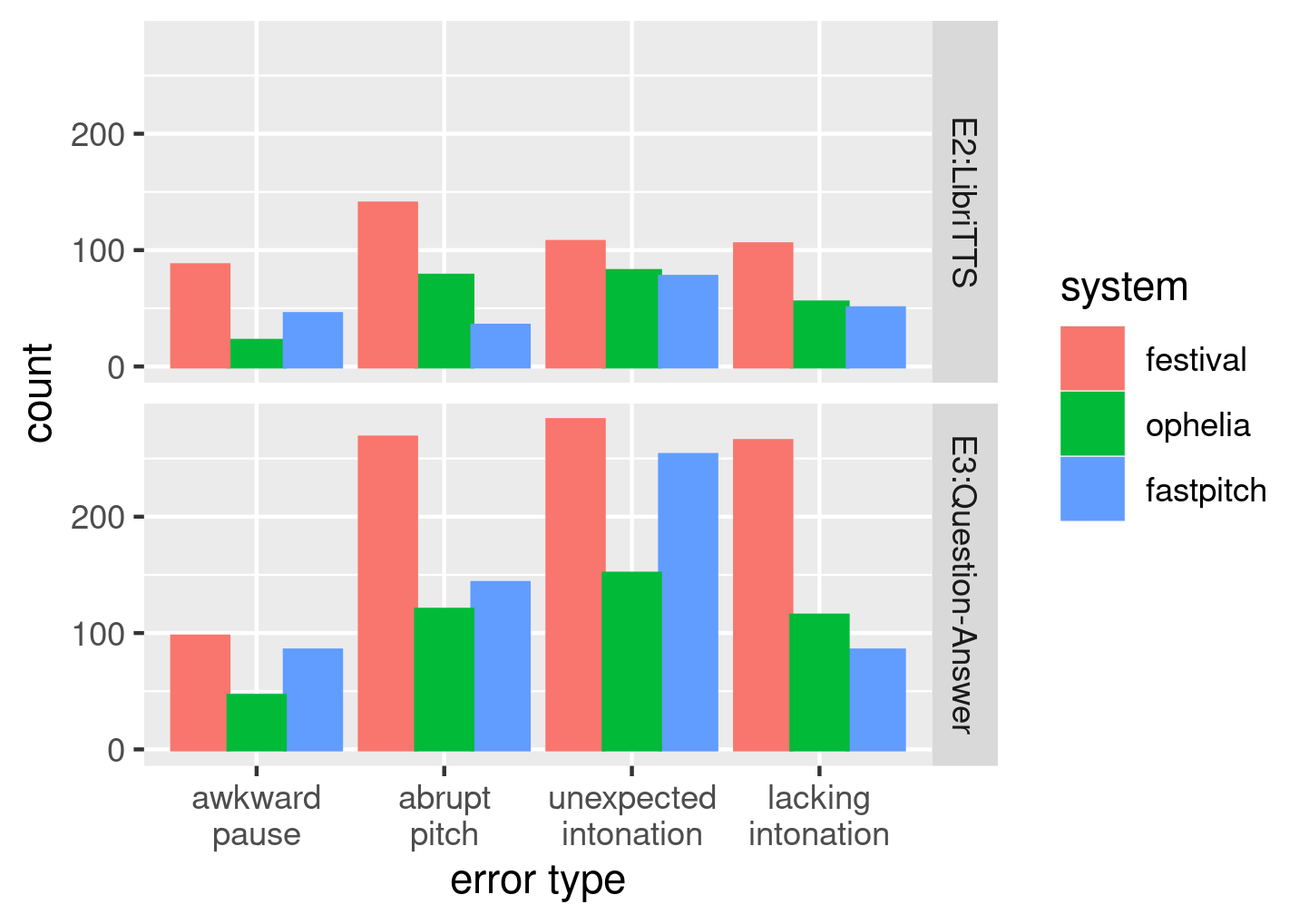}
    \caption{Counts of error types per system for E2, E3.}
    \label{fig:error-type-bar}
\end{figure}

Figure~\ref{fig:error-type-bar} shows the distribution of error types selected per experimental condition. This again supports the idea that prosodic expectations had a larger role when evaluating question-answer pairs.
Overall, we see a lower number of error types selected for the LibriTTS set than the question-answer set.  In particular, participants seemed less likely to detect abrupt pitch changes in FastPitch compared to Ophelia in the LibriTTS stimuli.  However, we see a marked increase in unexpected intonation errors for the FastPitch question-answer set. In contrast, FastPitch garnered slightly less `lacking intonation' errors than Ophelia for the question-answer stimuli.  This suggests that issues with FastPitch came from an excess of prosodic variation, which was more salient for the question-answer set.    

\begin{table}[t]
\centering
\begin{tabular}{llccc}
  \hline
 Test set & system & $\alpha$ & $\alpha_p$ & $N_p$ \\ 
  \hline
LibriTTS & festival & 0.12 & 0.18 & 7.67 \\ 
    & ophelia & 0.16 & 0.26 & 5.60 \\ 
    & fastpitch & \bf 0.24 & \bf 0.28 & \bf 4.90 \\ 

  \hline
Question-Answer  & festival & 0.10 &\bf  0.24 & 6.53 \\ 
    & ophelia & \bf 0.28 & 0.18 & \bf 3.90 \\ 
    & fastpitch & 0.18 & \bf 0.24 & 5.20 \\ 
   \hline
\end{tabular}
\caption{Mean interannotator agreement: Krippendorff's $\alpha$, Krippendorff's $\alpha$ restricted to participants that marked at least one error in the stimulus ($\alpha_p$),
the number of participants who marked an error in a stimulus ($N_p$, max 10).}
\label{tab:alpha}
\end{table}

We originally expected that 
the question-answer pairs would lead to greater interannotator agreement on error locations compared to the LibriTTS data due to stronger prosodic expectations. To investigate this, 
Krippendorff's $\alpha$ \cite{krippendorff2011computing} was calculated across the annotations for each stimulus in E2 and E3. We used the coding error=1, no error=0 for the annotation. 
Since participants could mark no errors in a stimulus, we added an additional `word' to each annotation marked 1 if the participant marked no other errors, and 0 otherwise.  This ensured that `no error' annotations would be counted as agreeing. To see more clearly if errors tended to be marked on the same words, we also calculated agreement per stimulus discarding annotations with no error marks ($\alpha_p$). We also count the number of participants who marked any error in a stimulus ($N_p$). 

Agreement statistics are shown in Table~\ref{tab:alpha}.  The LibriTTS results were as expected: as the lowest quality system, Festival, displays the lowest inter-annotator agreement, while Ophelia and FastPitch exhibit greater agreement for both $\alpha$ and $\alpha_p$. The mean values for $N_p$ also align with the PMOS ranking.  For the question-answer test set,  $\alpha$ and $N_p$ also reflects the PMOS ranking. However, $\alpha_p$, is higher for Festival and FastPitch, indicating that, while there was less agreement on whether there was an error in a stimuli: when participants marked an error they were more likely to choose the same word in the FastPitch and Festival cases. However,  we note that both types of $\alpha$ value are still in the low agreement range, so other types of errors likely came into play (cf.  Figure~\ref{fig:error-type-bar}).

\begin{figure}[t]
    \centering
    \includegraphics[scale=0.45]{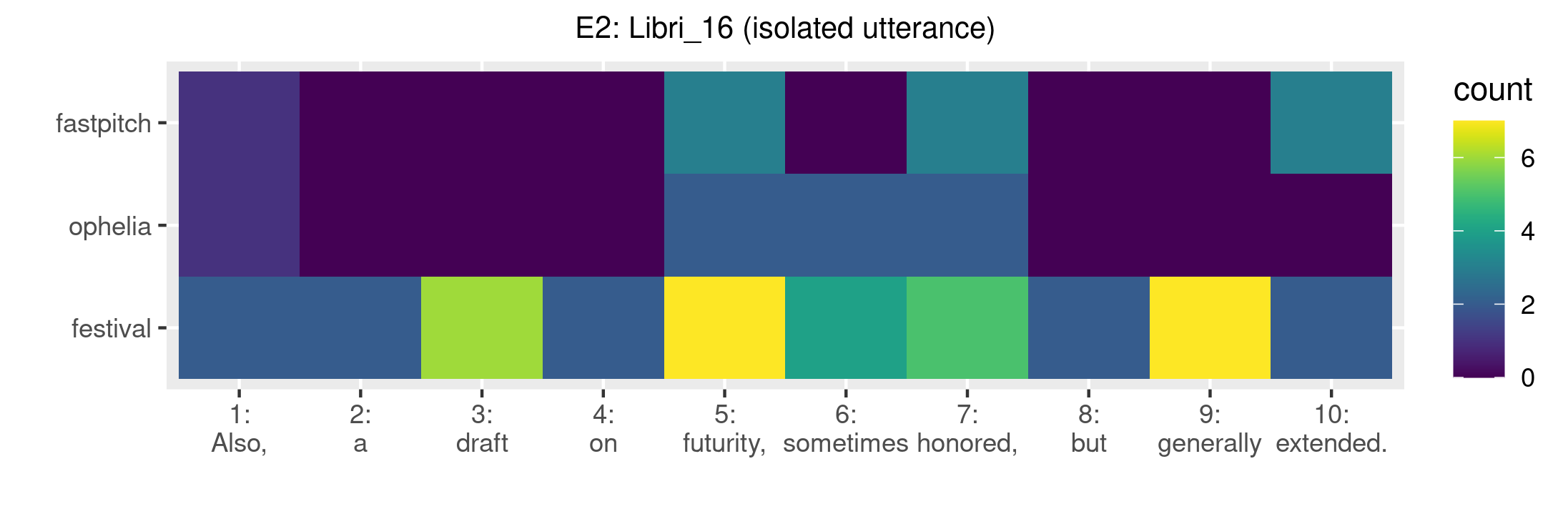}
    \caption{Error Heatmap (Libri16), PMOS:  Festival=1.7, Ophelia=3.6, Fastpitch=2.90.}
    \label{fig:libri16}
\end{figure}

\begin{table}[t]
\centering
\begin{tabular}{lcc}
  \hline
 System & E2: LibriTTS & E3: Question-Answer \\ 
  \hline
Festival & 0.50 & 0.52 \\ 
Ophelia & 0.67 & 0.45 \\ 
FastPitch & 0.73 & 0.60 \\ 

   \hline
\end{tabular}
\caption{Proportion of time the most error marked word in a stimulus preceded punctuation. Note, LibriTTS includes much more within utterances punctuation and punctuation variation than the Question-Answer set.}
\label{tab:punc}
\end{table}

A benefit of the error annotation is that we can visualize the distribution of errors across systems to direct further investigation. For example, Figure~\ref{fig:libri16} shows the error heatmap for a LibriTTS stimulus where FastPitch was rated lower than Ophelia in PMOS. This shows that error markings for FastPitch tended to occur on words attached to punctuation marks. To check whether this occurred more generally, we calculated the proportion of times that the most error marked word per stimulus preceded punctuation. The results in Table~\ref{tab:punc} indicate that punctuation was a more salient issue for FastPitch than for Ophelia.

\begin{figure}
    \centering
    \includegraphics[scale=0.1]{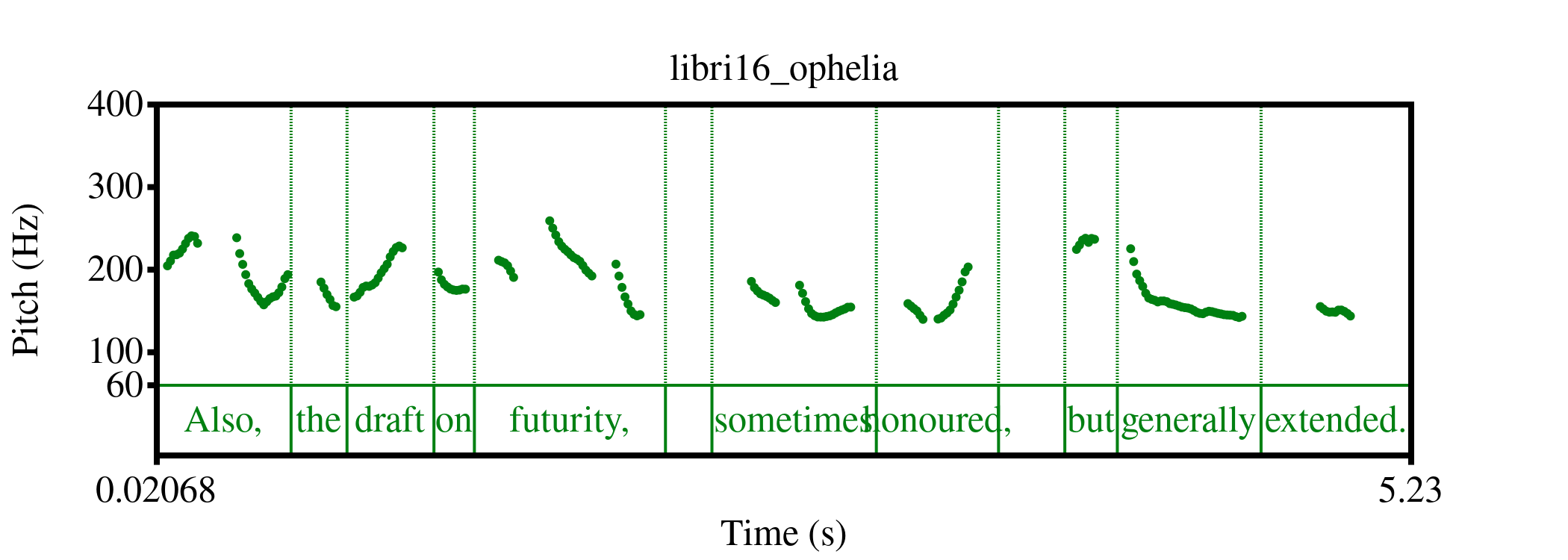}
    \includegraphics[scale=0.1]{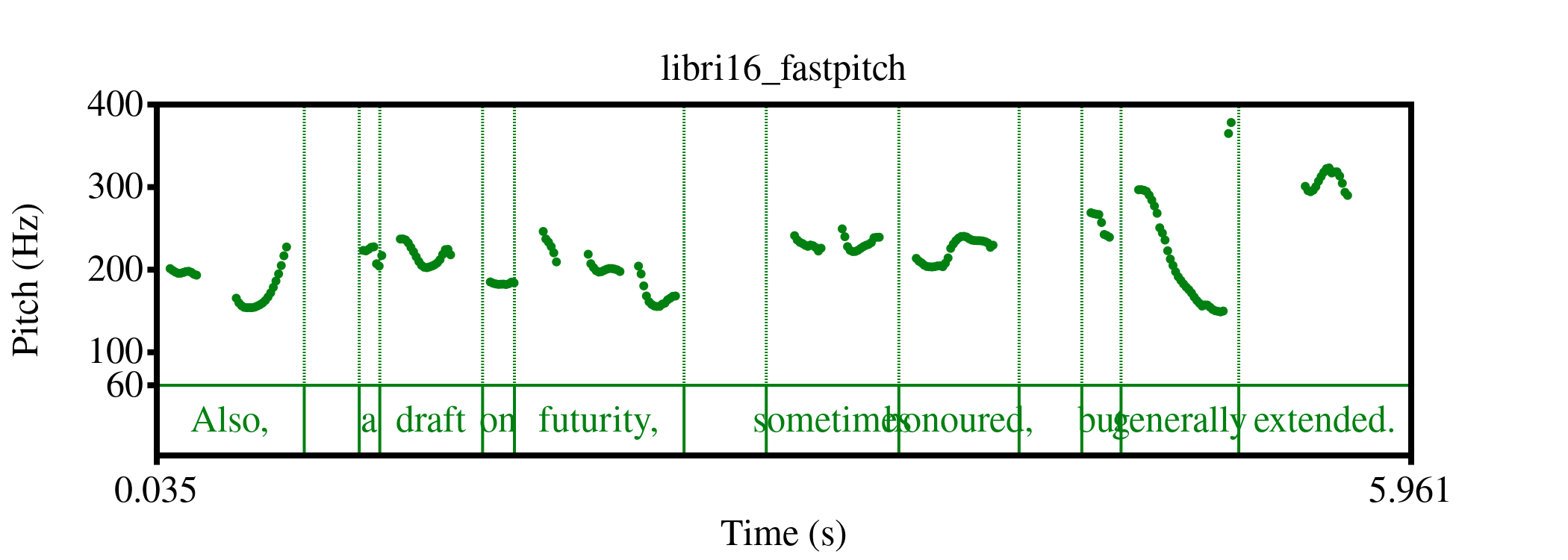}
    \caption{F0 differences for Libri16: FastPitch has unexpected pitch accents on before punctuation.}
    \label{fig:libri16-praat}
\end{figure}


Figure~\ref{fig:libri16-praat} shows F0 contours  corresponding to the heatmap in Figure~\ref{fig:libri16}.  Out of the 11 error types checked for the FastPitch version, 5 were for `awkward pause', 2 for `abrupt pitch' and 4 for `unexpected prosody', while for Ophelia 3/5 votes were for `lacking intonation'. On the FastPitch version we observe unexpected H* like pitch accents on `honoured,' and `extended.', while the Ophelia rendition has a continuation rise on `honoured,' and a fall to low pitch through `extended'. This supports the idea that punctuation produces specific prosodic expectations which were violated by the high level of prosodic variability (i.e., expressiveness) of FastPitch.  



\begin{figure}
    \centering
    \includegraphics[scale=0.45]{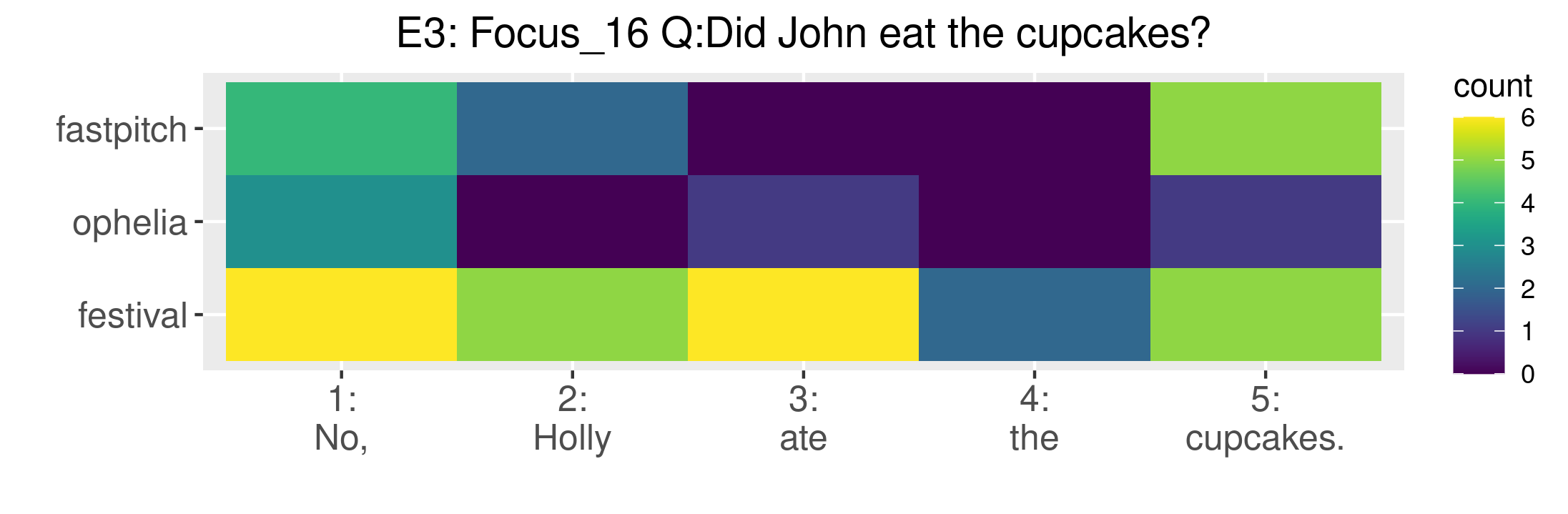}
    \caption{Error Heatmap (focus16); PMOS: Festival=1.5 Fastpitch=3.2 Ophelia=3.0}
    \label{fig:focus16}
        \vspace{-4mm}
\end{figure}

\begin{figure}[t]
    \centering
    \includegraphics[scale=0.11]{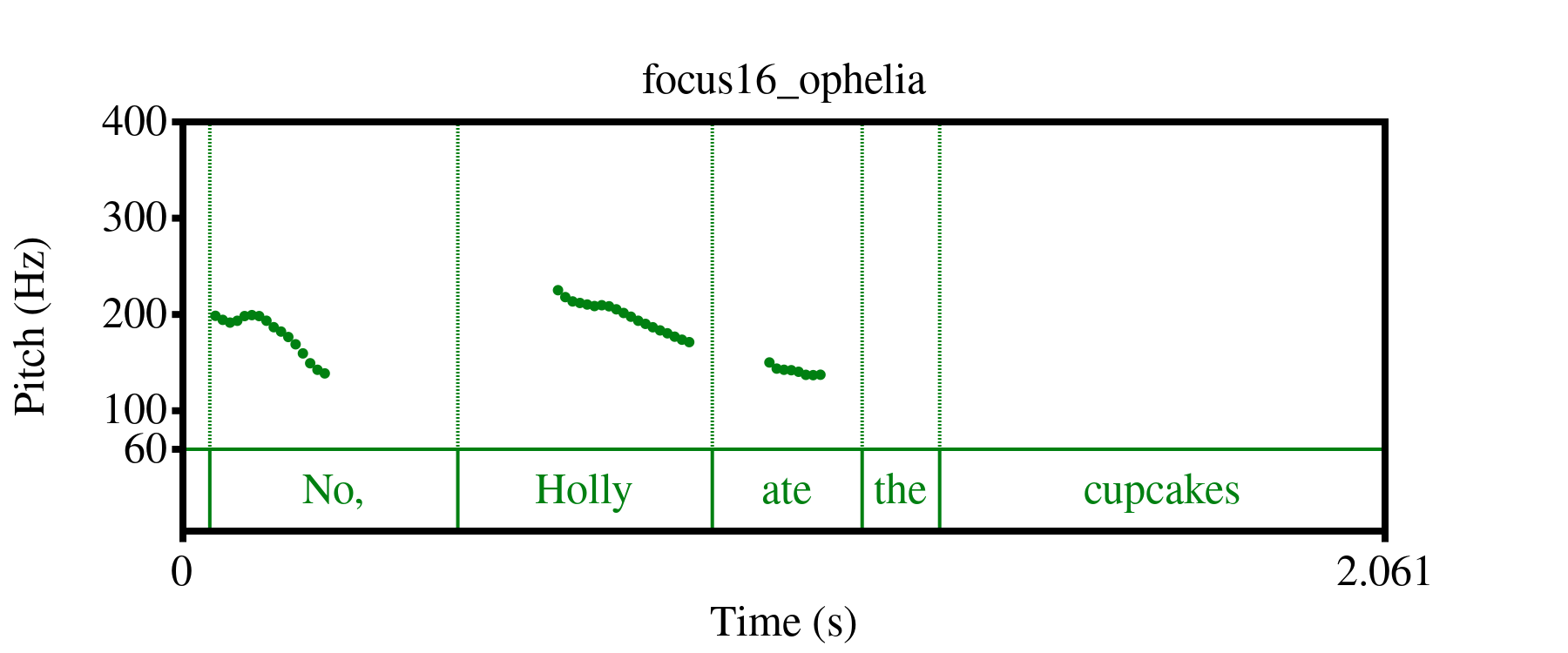}
    \includegraphics[scale=0.11]{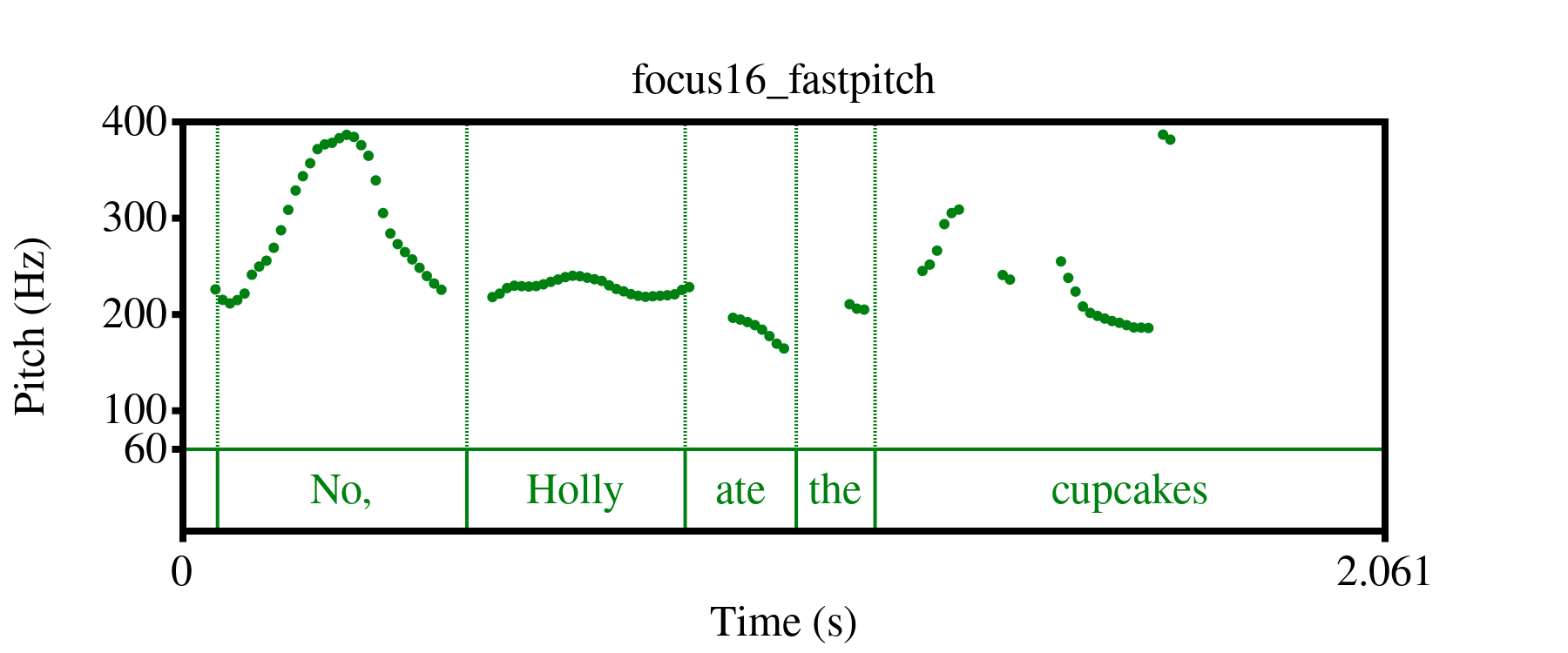}
    \caption{F0 differences (focus16): FastPitch produces an extra prominence on 'cupcakes' (cf. Figure~\ref{fig:focus16})}
    \label{fig:focus16-praat}
    \vspace{-4mm}
\end{figure}

Similarly, Figure~\ref{fig:focus16} shows error distributions for a contrastive focus example.  Figure~\ref{fig:focus16-praat} indicates the error marks on `cupcakes' in the FastPitch version are due to an unexpected pitch accent: `cupcakes' is given relative to the context question and so should be deaccented. Interestingly, pitch tracking for the Ophelia version fails on `cupcakes' due to issues in the signal quality, resulting in creaky-sounding (i.e., low pitched) voice. This is in line with information structure expectations but still may have reduced overall stimuli PMOS. We can also see that the large pitch excursion on the FastPitch `No' was perceived as an error.  While this doesn't produce an information structural clash, it does present an unexpected level of emphasis without further contextual information to justify it. 

\section{Discussion}

Our results indicate that error markings are consistent with rankings from MOS tests. 
However, differences between systems changed when participants were primed to focus on prosodic issues rather than naturalness in general. This means that the large lead FastPitch had over Ophelia in the naturalness (E1) is likely due to improvements in speech quality separate to prosody. It appears participants did separate out prosody and other quality issues, even for Festival which exhibited much lower naturalness than our neural TTS models. In fact, the error rate measure was better than PMOS at discriminating system prosody when combined with the question-answer test set, where prosodic expectations are more constrained. 

It's important to note that neither FastPitch or Ophelia take into account preceding context in their generation processes. The lower ranking of FastPitch in the question-answer test is likely due to overly-variable (unexpected) prosody, rather than Ophelia being intrinsically better in context. MOS scores for Ophelia were generally more variable, especially in E3. So, it is likely that Ophelia generates a more typical `reading style' intonation, which works well for some question-answer pairs, but not for others. 

Default `reading' intonation can work well for narrative-style (e.g., LibriTTS), but can be problematic when prosodic expectations are stronger, such as in task-oriented dialogues. This motivates more design and use of context sensitive stimuli, where factors contributing to prosodic expectations are well understood. A key factor for English prosody is information structure, but other factors will likely be important for other languages.   
In general, there are many paths which might lead listeners to give similar a (P)MOS, especially if the stimuli contains other types of errors (cf. relatively low interannotator agreement).  This indicates that simply asking more specific MOS questions isn't enough to pinpoint differences in systems.  The results of the current study support the case for more fine-grained error analysis. 


The real-time marking used in this study can help TTS researchers and designers understand what non-expert listeners pay attention to when they evaluate and perceive synthetic speech. For example, our study suggests that listeners have specific expectations about what should happen around prosodic boundaries signalled by punctuation. 
Similarly, the results from the question-answer testset provide evidence for an expectation-driven view of prosody perception \cite{Breen2010,roettger2019mapping}.
Further analysis of the acoustic properties around error marks will help improve our understanding of these expectations in future work.
Similarly, this method may shed light on cases where additional context actually allows for greater prosodic variability than in the isolated case \cite{clark2019ssw}.

While we have not done a comprehensive usability study of this method, 
many participants reported in the post-survey feedback form that they didn't find the experiment tiring and  were even entertained by the experiment. 
This feedback suggests that the methodology is feasible and may help mitigate loss of attention in evaluating long-form TTS \cite{govender2018using}. The fact that non-expert listeners can be used for the evaluation means that the methodology is scalable and gives a more realistic account of how a synthetic voice is perceived than a method using expert prosodic labelling. 

\section{Conclusion}

This study introduced a novel evaluation paradigm that augments the standard MOS test with an RPT-based error marking task. Our experiments showed how this fine-grained error marking can uncover differences in systems in prosody generation. 
We confirmed that our error marking method can be used to distinguish prosodic quality of different TTS systems with a greater degree of precision than MOS-only tests. The experiments highlighted the  usefulness of including question-answer test materials, and more generally stimuli which induce clear prosodic expectations. This new test set provided evidence for an expectation-driven model of prosody perception in TTS. This highlighted the fact that the high prosodic variability, often associated with  expressive TTS, may be perceived as errors when it doesn't match prosodic expectations induced by the context. 


Future work will involve a more detailed study of the acoustic properties of the error markings, and the priming effect of RPT-based error marking on PMOS scores. We would also like to extend this work to evaluate other long-form synthesis, e.g. narrative and conversational TTS, to better understand when contexts admits prosodic variation.  We would also like to extend the paradigm to evaluate, for example, speaker intent and  speaker stance.

\textbf{Acknowledgements.} This work was supported in part by: ANID, Becas Chile, nº 72190135.

\bibliographystyle{IEEEtran}

\bibliography{references,mybib,clai}

\end{document}